\documentclass[pra,showpacs,subeqn,graphicx]{revtex4}
\usepackage[dvips]{color}
\usepackage{newlfont}
\usepackage{amssymb}
\usepackage{amsfonts}
\usepackage{amsmath}
\usepackage{graphicx}
\usepackage{psfrag}
\usepackage{bm}

\newcommand{\beq}{\begin{equation}}
\newcommand{\eeq}{\end{equation}}

\newcommand{\matwo}[1]{\left(\begin{array}{cc} #1 \end{array}\right)}

\newcommand{\ket}[1]{|#1\rangle}

\begin{document}

\title{Analysis of a Magnetically Trapped Atom Clock}

\author{D. Kadio$^1$ and Y.B. Band$^{1,2}$}
\affiliation{$^1$Departments of Chemistry and Electro-Optics, and The 
Ilse Katz Center for Nano-Science, Ben-Gurion University of the Negev, 
Beer-Sheva 84105, Israel\\ 
$^{2}$Atomic Physics Division, A267 Physics, National Institute of 
Standards and Technology, Gaithersburg, MD 20899}

\begin{abstract}

We consider optimization of a rubidium atom clock that uses
magnetically trapped Bose condensed atoms in a highly elongated trap,
and determine the optimal conditions for minimum Allan variance of the
clock using microwave Ramsey fringe spectroscopy.  Elimination of
magnetic field shifts and collisional shifts are considered.  The
effects of spin-dipolar relaxation are addressed in the optimization
of the clock.  We find that for the interstate interaction strength
equal to or larger than the intrastate interaction strengths, a
modulational instability results in phase separation and symmetry
breaking of the two-component condensate composed of the ground and
excited hyperfine clock levels, and this mechanism limits the clock
accuracy.
\end{abstract}


\maketitle

\section{Introduction}

The most accurate atomic clocks in operation today are based either on 
trapped single ions or on atomic beams.  The advantage of ion clocks 
is that a single ion can be trapped very tightly by static electric 
fields such that optical transitions do not cause significant heating 
or an escape of the ion.  These optical transitions are decoupled from 
the trapping potential such that both ground and excited atomic states 
feel the same potential.  On the other hand, one advantage of atomic 
beam clocks is the large number of atoms in a beam, such that the 
quantum projection noise can be decreased by 2-3 orders of magnitude 
with respect to the single ion clock.  A clock based on a thermal 
atomic beam suffers from the velocity distribution that limits the 
transition bandwidth.  Using a Bose-Einstein condensate (BEC) would 
significantly ameliorate this problem.  A new type of atomic clock 
based on neutral atoms trapped in a deep ``magic-wavelength'' optical 
lattice (magic because the transition does not have an optical light 
shift because the difference between the ac polarizabilities vanishes 
at the wavelength of the optical lattice) has recently been suggested 
\cite{Takamoto_03, Katori_03}.  This kind of clock can be operated on 
an optical transition, rather than a microwave transition, and 
promises to be most accurate, but clocks of this type have not yet 
been fully characterized.  

Here we consider whether a good atomic clock can be based on a more
common type of trapped ultracold atom configuration, i.e., on a BEC in
a magnetic trap.  Trapped BECs can have many atoms, which gives them
the large number advantage mentioned above.  This kind of clock can be
much more accurate than a thermal cloud clock because the Doppler
effect in a thermal clould can severely limit clock performance (see
Sec.~\ref{dynamics} below).  This effect is negligible for a BEC. In
addition a BEC cloud has a well defined energy determined by the
chemical potential that is uniform over the BEC and this helps in
lowering the variance of the clock frequency.  This kind of clock
might be miniaturized, as microtraps for atomic BECs can be created
above a fabricated chip.  As has now been fully demonstrated, magnetic
microchip traps can be used to manipulate neutral atoms on the
micrometer scale \cite{Folman_02}.  A high density, coherent atom
source can be created via Bose-Einstein condensation on an atom chip
\cite{chip_BEC}, and ``atomic conveyor belts'', waveguides, and beam
splitters can be implemented on atom chips \cite{Folman_02}.  It is
therefore intriguing to entertain the possibility of creating an
atomic clock on an atom chip \cite{Treutlein_02,Knappe_05}.  Hence, it
is important to study theoretically and experimentally the potential
of this kind of clock.  One experiment of this kind, using Ramsey
spectroscopy \cite{Ramsey_50}, has already been carried out
\cite{Treutlein_02}, and another experiment has been performed in a
macroscopic magnetic trap \cite{Harber_02}, but using the same
spectroscopic method.  More recent mesoscopic atom clocks using
coherent population trapping have been reported \cite{Knappe_05}, and
have the benefit of allowing compact optical light sources.

Specifically, we consider a BEC in a magnetic trap and investigate a
clock based on the Ramsey separated field spectroscopy method
\cite{Ramsey_50} in a highly elongated trap.  The quasi-1D geometry of
an elongated trap has the advantage of further reducing the inelastic
ultra-cold collisions as shown in Ref.~\cite{Yurovsky_06}.  As in
Refs.~\cite{Treutlein_02} and \cite{Harber_02}, we consider a
two-photon microwave transition between two $^{87}$Rb hyperfine states
with an atomic frequency $\nu_0 \simeq 6.8$ GHz.  We treat the
dynamics of the clock in mean-field and consider the amplitude and
phase of the order parameters for the ground and excited clock states
of the system, solving the coupled set of 1D Gross-Pitaevskii
equations to analyze the microwave clock frequency shift due to
collisional and magnetic field effects.  We determine the clock
frequency shift introduced by the external magnetic potential and the
kinetic energy of the Bose condensed gas, both of which are influenced
by the difference in the size of the two atomic wave packets.  The
clock is designed to run with $^{87}$Rb atoms in a magnetic field
regime where the two hyperfine levels correlating with $5^{1}S_{1/2}
\ket{f=1,m_f=-1}$ and $5^{1}S_{1/2} \ket{f=2,m_f=1}$ experience the
same first order Zeeman shift \cite{Treutlein_02, Harber_02,good_q_n}.
The collisional frequency shift from the resonance frequency $\nu_0$
can be calculated {\it \'{a} la} Ref.~\cite{Harber_02}.  As we shall
see, the collisional shift can be cancelled by using the Zeeman shift
\cite{Harber_02} and by optimizing the population difference in the
ground and excited states \cite{Gibble_95}.  The latter is possible
for an interstate interaction strength larger or smaller than both the
ground and excited intrastate interaction strength.  The clock
run-time is limited by atom loss due to collisional spin dipolar
collisional relaxation of the excited state \cite{Harber_02}.  For a
$^{87}$Rb condensate at high density ($12.6 \times 10^{13}$
cm$^{-3}$), the collisional dipolar loss in the excited state has been
experimentally measured \cite{Harber_02}.  Atom loss led to a total
density to drop by 3\% in $20$ ms.  Nevertheless it is important to
have a reasonable atomic number density $n$ to compensate the effects
of quantum fluctuations; the uncertainty, as quantified by the Allan
standard deviation $\sigma$ scales as $n^{-1/2}$.  So an optimization
of the density of atoms is necessary to reduce the quantum
fluctuations and the collisional dipolar relaxation in order to
increase the clock time.  The first experiment with this type of clock
using a trapped thermal cloud of $^{87}$Rb atoms containing about
$1.5\times 10^{4}$ atoms with a density of atoms less than $5\times
10^{12}$ cm$^{-3}$, yielded an Allan standard deviation
\begin{equation}
\sigma(\tau) = 1.7\times 10^{-11} \tau^{-1/2} \, \,
{\mathrm{Hz}}^{-1/2} \,,
\end{equation} 
where $\tau$ is the averaging time \cite{Treutlein_02}.  For the trap
parameters used here, i.e., with radial frequency $\omega_r/2\pi =
120$ Hz and axial frequency $\omega_z/2\pi=0.5$ Hz, the Allan standard
deviation is of order of magnitude $10^{-12}\sqrt{T_c/\tau}$, where
$T_c$ is the cycle period.  This trap confines the geometry to
quasi-1D and has the advantage of further reducing the collisional
dipolar relaxation \cite{Yurovsky_06}.

Some additional crucial limitations might make a magnetically trapped
BEC unsuitable.  A significant limitation is the collisional
interaction between the atoms.  Particularly problematic is the
difference in the $s$-wave scattering length between atoms occupying
different hyperfine levels which affects the collisional shift of the
clock frequency.  The collisional shifts of rubidium atoms are
relatively small compared to cesium atoms \cite{Fertig_00,
Sortais_00}, for example, but they can still be significant if many
atoms are tightly trapped together.  In order to minimize collisional
shifts, we shall employ a method of overcoming collisional shifts by
adjusting the ground to excited state ratio during the Ramsey fringe
spectroscopy.  Moreover, the run-time of the clock is also limited by
the dynamics of the atomic cloud that can result in phase separation
of the two spin components \cite{Hall_98,Ho_96,Pu_98,
Timmermans_98,Tripp_00}.  We find that a {\em modulational
instability} results in the dynamics and the evolution depends on both
the density of atoms and the balance between the interstate and
intrastate interaction strengths.  This gives rise to phase separation
and symmetry breaking of the two-component condensate for the ground
and excited clock levels that occurs after the first $\pi/2$ Ramsey
pulse that puts the atoms in a superposition of the ground and excited
state.  The modulation instability limits the clock accuracy.  It
therefore appears that magnetically trapped BEC clocks on an atom chip
cannot promise to be the most accurate type of clock.  The most
significant limitation to the clock stability arises from the dynamics
of the atomic cloud that creates a phase separation of the two wave
packets for the ground and excited state.  The time dependence of the
phase separation depends on the density of atoms and on the interstate
interaction strength; the smaller the density and/or the smaller the
interstate interaction strength, the longer the phase separation time.
Hence, a very weak axial trapping frequency (e.g., $\omega_z/2\pi <
0.5$ Hz) resulting in a lower density of the atoms, allows an
increased interrogation time and/or a greater total number of atoms,
and therefore a further increase the stability of the clock beyond
$10^{-12} \sqrt{T_c/\tau}$.

The paper is organized as follows.  The model of the clock based on 
Ramsey spectroscopy is described in Sec.~\ref{M_BEC_clock}.  Section 
\ref{numerical_method} briefly presents the numerical approach we use 
to analyze the clock.  In Sec.~\ref{dynamics} we depict the quasi-1D 
dynamics in a trap that is very tight in two directions, and describe 
why the spin-relaxation collision mechanism, as well as other 
inelastic scattering processes, is suppressed in a 1D geometry. 
Sec.~\ref{NR} describes the results obtained by numerically solving 
the coupled Gross-Pitaevskii equations for the order parameters of the 
ground and excited clock states.  In Sec.~\ref{improvement} we discuss 
two ways to improve and optimize the stability and accuracy of the 
clock by cancelling the collisional shift.  Section \ref{Conclusion} 
concludes the paper.

\section{Microwave BEC Magnetic Clock Using Ramsey Fringes} \label{M_BEC_clock}

We consider an atomic BEC trapped in an external magnetic potential.  
The spatial variation is harmonic about the trap minimum.  The atoms 
are initially in the ground electronic state, labeled $\ket{1}$, and a 
radio frequency field can transfer atoms into an excited state labeled 
$\ket{2}$.  More specifically the levels $|f,m> = |2,1>$ and $|1,-1>$ 
are used, and the transition involves a combination of a microwave 
pulse at 6.8 GHz to transfer the atoms from $|2,1>$ to $|1,0>$ and 
then another RF pulse to transfer them from $|1,0>$ to $|1,-1>$.  The 
$|1,-1>$ state is trapped with the same potential as $|2,1>$ if the 
magnetic field at the trap bottom is around 3.23 G \cite{good_q_n}.  

The clock described here uses the Ramsey separated field method
\cite{Ramsey_50}.  The atomic cloud interacts with two short microwave
pulses separated by a time $T$; each pulse has pulse area close to
$\pi/2$.  A spatial inhomogeneity of the atomic energy levels is due
to the spatially dependent Zeeman energy due to the magnetic field
varying with position.  Clearly, this can adversely affect the clock
frequency.  This effect is minimized by using a pair of energy levels
which experience the same trapping potential at a particular magnetic
field strength.  Refs.~\cite{Treutlein_02} and \cite{Harber_02} showed
that at a magnetic field of $\sim 3.23$ G, the $\ket{1} \equiv
\ket{f=1,m_f=-1}$ and $\ket{2}\equiv \ket{f=2,m_f=1}$ hyperfine levels
of the $5S_{1/2}$ ground state of $^{87}$Rb experience the same first
order Zeeman shift such that the differential shift of the two levels
across the cloud was $\sim 1$ Hz.  The collisional shift also
contributes to the spatial inhomogeneity of the atomic transition
energy level across the cloud since the density of the cloud varies
with position.  However, as noted in Ref.~\cite{Harber_02}, it may be
possible to cancel the Zeeman shift with the collisional shift.  The
stability and accuracy of the clock are further improved and optimized
by cancelling the collisional shift (as we shall see in
Sec.~\ref{improvement}).

The initial condensate starts in the ground state $\ket{1}$, and after 
the first $\pi/2$ pulse, which we model by Bloch sphere dynamics 
\cite{Vanier_89} assuming that the pulse duration $\tau_p$ is 
extremely fast compared to other time-scales, we solve a set of 
coupled Gross-Pitaevskii equations to describe the dynamics of the 
two-component ($\ket{1}$ and $\ket{2}$) wave packets.  After a time 
$T$ a second short $\pi/2$ pulse is applied.  For an intense short 
near-resonant pulse, the solutions of the optical Bloch equations for 
a two level atom gives the following unitary transformation operator: 
\begin{equation} \label{Rabi_trans}
\mathbf{U}(t)=\matwo{\cos(\Omega_g t/2) -
i\frac{\Delta \nu}{\Omega_g}\sin(\Omega_g t/2) &
-i\frac{\Omega}{\Omega_g} \sin(\Omega_g t/2) \\ 
-i\frac{\Omega}{\Omega_g}\sin(\Omega_g t/2) & \cos(\Omega_g t/2)+
i\frac{\Delta \nu}{\Omega_g}\sin(\Omega_g t/2)} \,, 
\end{equation}

where $\Omega$ is the Rabi frequency, $\Delta \nu$ is the detuning and 
$\Omega_g=\sqrt{|\Omega|^2+\Delta \nu^2}$ is the generalized Rabi 
frequency.  For example, if an atom is initially in state $\ket{1}$ 
and interacts with an on-resonant $\pi/2$ pulse, it evolves to the 
state $(\ket{1}+i\ket{2})/\sqrt{2}$.  This transformation can be used 
to describe the effects of both the first and second Ramsey pulses.  

\section{Mean-field Analysis of Clock Dynamics}  

The performance of the clock is affected by the dynamics of the
two-component BEC after the first $\pi/2$ Ramsey pulse.  We shall see
below that, because of the crossed interaction energy of the two spin
components created after the first $\pi/2$ Ramsey pulse, the system
becomes unstable, and the components eventually undergo a local phase
separation that leads to symmetry breaking.  The phase separation of
the spin components limits the Ramsey interrogation time and hence the
stability of the clock.

In this section, we first describe the numerical methods used to 
investigate the clock dynamics in mean-field.  Many-body effects can 
also be included as formulated in Ref.~\cite{Band_06}, but we shall 
not do so here.  Then we discuss the advantage of operating the clock 
in a highly elongated trap configuration.  We present numerical 
results for this configuration and analyze them.

\subsection{Numerical method}  \label{numerical_method} 

We investigate the clock dynamics in mean-field.  The initial zero 
temperature condensate wave function (order parameter) is obtained by 
numerically determining the lowest eigenstate $\psi({\mathbf r})$ of 
the time-dependent Gross-Pitaevskii equation for particles of mass m, 
confined in an external potential $V_{\mathrm{ext}}({\mathbf r})$ and 
a mean-field interaction energy due to contact two-body interactions 
with coupling strength $g_{11}={4\pi \hbar^{2}a_{11}}/{m}$ where 
$a_{11}$ is the $s$-wave scattering length for atoms in the ground 
state.  This is accomplished with an imaginary time split-step Fourier 
transform method.  The effect of the first pulse that couples the two 
atomic spin states is modeled using a unitary transformation on the 
zero temperature ground state wave function and gives two wave 
functions representing the ground state and the excited state atoms: 
\begin{equation} 
\psi_{i}({\mathbf r},0) = A_{i} \psi({\mathbf r}) 
\end{equation} 
where $i=1,2$ correspond to ground and excited state labels 
respectively, and $A_i$ is the complex amplitude of state $i$ obtained 
using Eq.~(\ref{Rabi_trans}).  We take the normalization of the 
initial condensate wave function such that $\int \big|\psi(r,0) 
\big|^{2}d{\mathbf r} = N$, where $N$ is the total number of atoms, 
and the amplitudes $A_i$ are determined by the Bloch sphere dynamics 
for the two levels in the presence of the microwave field inducing the 
transition \cite{Vanier_89}.  The amplitudes $A_i$ satisfy $0 \leq 
\big|A_{i}\big|^2 \leq 1$ and $\sum_{i=1}^{2} \big|A_{i}\big|^2 = 1$.  
The two component condensates evolve according to 
\begin{equation}
\begin{array}{rcl}
i\hbar \, \frac{\partial \psi_i({\mathbf r},t)}{\partial t}&=&\Bigg(
-\frac{\hbar^{2}\bigtriangledown^{2}}{2m} +V_{\mathrm{ext}}({\mathbf 
r})+(-1)^{i}\frac{\hbar\omega_{21}}{2} +\sum_{j=1,2}
g_{ij}\big|\psi_{j}({\mathbf r},t)\big|^2 \Bigg) \psi_{i}({\mathbf 
r},t)
\end{array} 
\end{equation} 
where the atomic resonance transition frequency is denoted as
$\omega_{21}$.  The interaction strength $g_{ij} = 4\pi \hbar^{2}
\alpha^{(2)}_{ij} a_{ij}/m$, with $i,j=1,2$, is defined in terms of the
$s$-wave scattering length for particles in states $i$ and $j$,
$a_{ij}$, and the two-particle correlation parameter for particles in
states $i$ and $j$ at zero separation between particles,
$\alpha^{(2)}_{ij}$ \cite{Harber_02,Band_00,Ketterle_97,Zwierlein_03,
Kheruntsyan_03,Naraschewski_99}.  The latter quantity is often denoted
as $g^{(2)}_{ij}$.  The values of the two-particle correlation parameter
$\alpha^{(2)}_{ij}$ is such that $0 \leq \alpha^{(2)}_{ii} \leq 2$ and
$1 \leq \alpha^{(2)}_{ij,i \neq j} \leq 2$ for Bosons.  For a
condensate, $\alpha^{(2)}_{ii} = 1$.  For the inter-state
(two-component) two-particle correlation at zero separation,
$\alpha^{(2)}_{12}$ in a condensate we considered two values,
$\alpha^{(2)}_{12} = 1$ and $\alpha^{(2)}_{12} = 2$.  We have learned
recently that $\alpha^{(2)}_{12}$ was measured to be nearly unity for
$^{87}$Rb \cite{Cornell}, but to we shall present results of
calculations for both values.

We propagate the two BEC components for a time $T$ between the Ramsey 
pulses by solving the coupled time-dependent Gross-Pitaevskii 
equations using the split step Fourier transform method.  During the 
propagation for a time $T$, the phase and amplitude of the component 
wave functions evolve with time.  After the time $T$, we apply again 
the unitary transformation operator corresponding to a $\pi/2$ pulse.  
We then integrate the component wave functions over space to obtain 
the probabilities for finding the atoms in the two states.  

\subsection{Dynamics of the clock in a highly elongated trap} \label{dynamics}

The effects of the mean-field collisional dipolar relaxation are 
important to investigate in order to optimize the accuracy and the 
stability of the clock.  As we shall see, phase separation of the two 
spin components due to mean-field dynamics, and loss of excited state 
atoms due to the collisional dipolar relaxation between atoms in the 
excited state are two factors that can significantly reduce the 
performance of the clock.  The collisional dipolar loss of the excited 
state for a $^{87}$Rb condensate at high density ($12.6 \times 
10^{13}$cm$^{-3}$) has been experimentally measured \cite{Harber_02}.  
Atom loss caused the total density to drop by 3\% in $20$ ms.  In 
order to reduce this effect, we propose to run the clock in a highly 
elongated trap; in a quasi-one-dimensional condensate the spin-dipole 
relaxation collisional loss can be made much smaller than in 3D 
\cite{Yurovsky_06}.  The inelastic rate coefficient for going from the 
incident channel $\beta$ to a final inelastic channel $\beta^{'}$ for 
strong confinement $K_{\text{conf},\beta^{'} \beta}$ is related to the 
3D inelastic rate $K_{\text{free},\beta^{'} \beta}$ by \cite{Yurovsky_06}, 

\begin{equation} 
K_{\text{conf},\beta^{'} \beta} = {a^{4}_{\perp} p^{2}_{0}\over 
4\hbar^2|a_{\text{eff}}|^{2}}K_{\text{free},\beta^{'} \beta} = 
\frac{2 \pi a_{\perp }^4 E_0}{\hbar^2} \sum_{\beta^{'}}\frac{|U_{\beta^{'}
\beta}|^2}{p_{\beta^{'}}} \,.  
\end{equation}

Here $p_{0}$ is the relative collision momentum and $E_0 = 
p^{2}_{0}/2\mu_r$ is the relative collision energy in the incident 
channel, $p_{\beta^{'}} = \sqrt{ 2\mu_r(E_0 + E_{\beta^{'}\beta})}$ is 
the relative momentum in the final inelastic $\beta^{'}$ channel, 
where $E_{\beta^{'}\beta}$ is the asymptotic energy difference of the 
two channels, $\mu_r= m/2$ is the reduced mass, and the parameter 
$U_{\beta^{'} \beta}$ is the coupling between the channels  $\beta^{'}$ 
and $\beta$.  This quasi-1D rate is reduced by a factor of 
${a^{4}_{\perp} p^{2}_{0}\over 4\hbar^2|a_{\text{eff}}|^{2}}$ from the 
3D rate, and is small for ultra-cold collision energies.  For high 
anisotropy of the trap, the 1D interaction strength is given by 
\begin{equation}
g_{ij}^{1D}= \frac{2\hbar^{2}\alpha^{(2)}_{ij} a_{ij}}{m a_{\perp}^{2}} \,, 
\end{equation}
where $a_{\perp}=\sqrt{\hbar/m \omega_r}$ is the radial harmonic 
oscillator length and $\omega_r/2\pi$ the radial trap frequency 
\cite{Olshanii_98}.  For sufficiently large $\omega_r$, radial profile 
is harmonic oscillator like, and the motion of the atoms are frozen in 
the radial direction.  We assume that the magnetic field is such that 
the first order Zeeman shift is the same for the two atomic internal 
states.  Now, if the frequency $\omega_z/2\pi$ is also made small, so the 
1D density is small, the nonlinear interaction term $g_{ij}^{1D}  
n$ can be made very small.  

Using a magnetically trapped thermal cloud can significantly reduce
the mean-field collisional shift compared to a BEC, however, a thermal
cloud has a Doppler width that increases the bandwidth of the clock
transition and can therefore limit the clock accuracy.  Indeed the
clock bandwidth at half maximum is in our case inversely proportional
to the interrogation time $T$; the larger the interrogation time, the
smaller the transition bandwith of the clock, and the Doppler effect
on the bandwidth becomes more significant.  With our interrogation
time (~ 0.5 s), the Doppler width must be much smaller than 1 Hz to be
negligible.  Hence, the temperature of the thermal cloud must be well
below 2.5 $\mu$K so as to be competitive with a BEC. For a 500 nK
$^{87}$Rb thermal cloud (i.e., above condensation) the Doppler width
is 0.4 Hz which is smaller than the bandwith of the clock transition.
Moreover, in a BEC the energy is determined by the chemical potential
that is uniform across the BEC and very well specified; this helps in
lowering the variance of the clock frequency.  The collisional
frequency shift calculated with $10^4$ atoms in an highly elongated
BEC corresponding to a density of $3 \times 10^{14}$ atoms/cm$^{3}$
for our trap geometry, is about 3 Hz, but the uncertainty in this
collisional shift (due to the uncertainty in the number of atoms and
the uncertainty in the value of the scattering length) is very much
smaller than 3 Hz, and hence the collisional shift can be largely
compensated for as far as the clock frequency is concerned.  Although
the collisional shift for a 500 nK thermal cloud is at least 100 times
smaller than that for the BEC, the Doppler width of the thermal cloud
is 0.4 Hz, and this is presumably much larger than the {\em
uncertainty} in the collisional shift of the BEC.

\subsection{Numerical Results} \label{NR} 

Our numerical calculations have been carried out with an axial
($\nu_z$) and radial ($\nu_r$) trap frequency of $0.5$ Hz and $120$ Hz
respectively, so the anisotropy ratio $\lambda \equiv \nu_r/\nu_z =
240$.  The three scattering lengths are taken to be $a_{11} =
100.44\,a_0$, $a_{22} = 95.47\,a_0$ and $a_{12} = 98.09\,a_0$
\cite{Harber_02}, where $a_0$ is the Bohr radius.  In order to better
understand the effect of the modulational instability on the clock, we
carry out the calculation with $\alpha^{(2)}_{11,22}=1$ and
$\alpha^{(2)}_{12}=2$.  The optimization of the number of atoms for
the frequencies given above that gives the best Allan deviation is of
the order of magnitude $10^4$ atoms.  The two pulses used for the
Ramsey separated field method are taken to be $\pi/2$ pulses, {\it
i.e.}, $\big|A_1\big|^2 = \big|A_2\big|^2 = 1/2$.

\begin{figure}[t]
\begin{center} 
\includegraphics[scale=1.2]{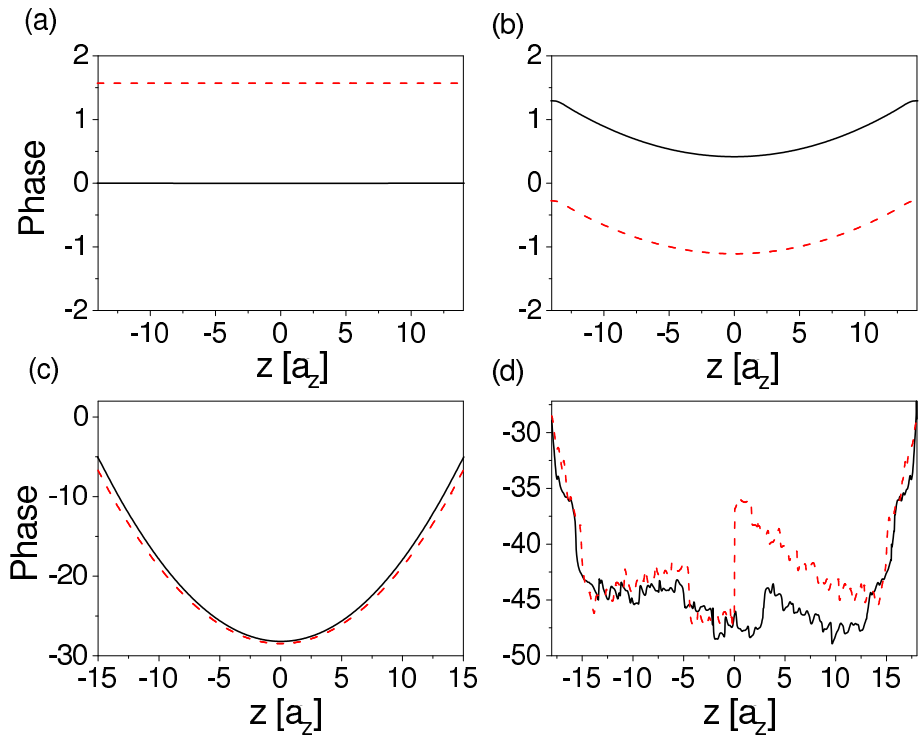} 
\caption{(color online) Phase of condensate ground state (solid curve)
and excited state (dashed curve) components for the interstate
two-particle correlation parameter at zero separation
$\alpha^{(2)}_{12} = 2$, after the first $\pi/2$ pulse as a function
of position, $z$, in the magnetic trap at times (a) $t=0$ ms, (b)
$t=6.3$ ms, (c) $t=180$ ms and (d) $t=500$ ms.  $a_{z} =
\sqrt{\hbar/m\omega_{z}}$ is the axial harmonic oscillator length.}
\label{phase} 
\end{center} 
\end{figure}

Fig.~\ref{phase} shows the phase $\theta_{i}(z,t)$ of the $i^{th}$
condensate wave function, $\psi_{i}(z,t)=|\psi_{i}(z,t)|$
exp$[i\theta_{i}(z,t)]$, as a function of position, $z$, in the
magnetic trap for $\alpha^{(2)}_{12}=2$.  Immediately after the first
$\pi/2$ pulse, the phase of the two spin components is spatially
uniform and their difference is $\pi/2$ (Fig.~\ref{phase} (a)) as is
easily understood from the transformation in Eq.~(\ref{Rabi_trans}).
Following the $\pi/2$ pulse, mean-field effects begin to create a
spatially varying phase across the two condensate wave packets
(Fig.~\ref{phase} (b) and (c)) \cite{Band_02}.  Beyond $t = 0.24$ s,
the spatially dependent variations in the phase appears completely
chaotic; the mean-field treatment has not only reached the point of
numerically limited accuracy but has actually lost its regime of
validity.

\begin{figure}[t] 
\begin{center} 
\includegraphics[scale=1.2]{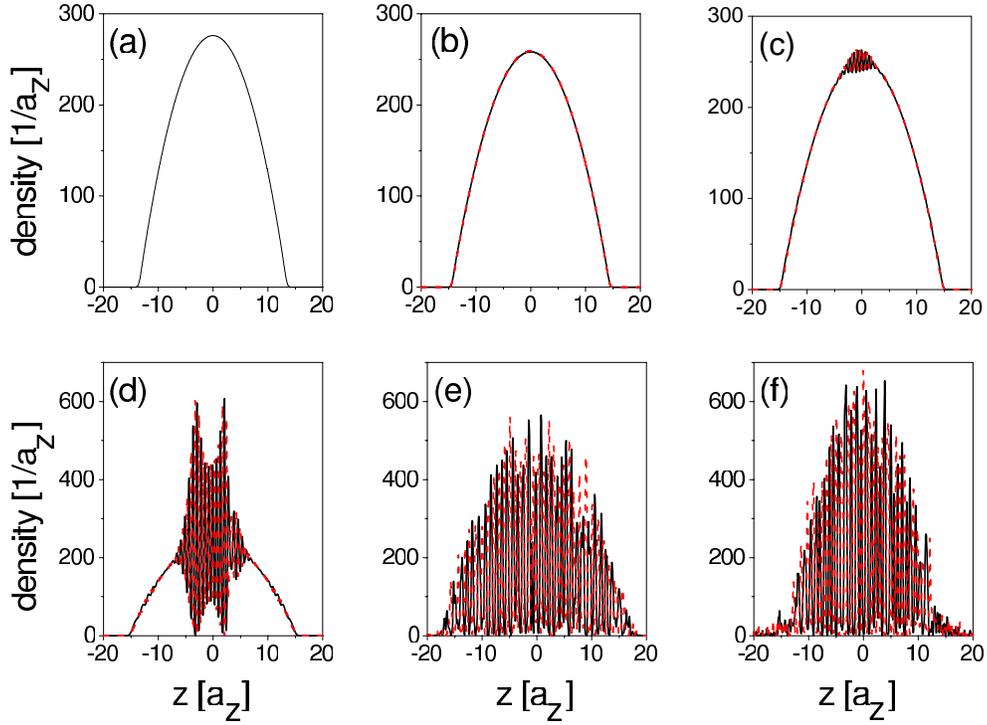} 
\caption{(color online) Condensate density of ground state (solid
curve) and excited state (dashed curve) components for the interstate
two-particle correlation parameter at zero separation
$\alpha^{(2)}_{12} = 2$, after the first $\pi/2$ pulse for a sequence
of inter-pulse times (a) $T=0$ s, (b) $T=0.18$ s, (c) $T=0.22$ s, (d)
$T=0.26$ s, (e) $T=0.5$ s and (f) $T=1$ s.  At $T=0$ s the two spin
components have the same amplitude since the microwave pulse is very
short but dephased of $\pi/2$ for a $\pi/2$ pulse.  $a_{z} =
\sqrt{\hbar/m\omega_{z}}$ is the axial harmonic oscillator length.}
\label{density_position} 
\end{center} 
\end{figure}

Fig.~\ref{density_position} shows the evolution of the position
dependent density of the two atomic states for a sequence of
inter-pulse times $T=0$ s (Fig.~\ref{density_position}(a)), $T=0.18$ s
(Fig.~\ref{density_position}(b)), $T=0.22$ s
(Fig.~\ref{density_position}(c)), $T=0.26$ s
(Fig.~\ref{density_position}(d)), $T=0.5$ s
(Fig.~\ref{density_position}(e)), and $T=1$ s
(Fig.~\ref{density_position}(f)), and $\alpha^{(2)}_{12}=2$.  The
density of the two components are almost identical at $T=0$ s and
$T=0.18$ s and are smoothly varying with position.  The density
profiles begins becoming irregular at the center of the condensates at
about $T=0.20$ s.  At later times a spiked structure whose amplitude
increases with time develops and some local phase separation occurs
(Fig.~\ref{density_position}(c)-(f)) due to the repulsive interaction
between the wave packet components.  A similar spiked structure of the
density as a function of position has been obtained numerically in the
regime of strong excitation of the BEC loaded in a 1D optical lattice
plus an asymmetric external magnetic trap by instantaneously giving a
large displacement to the initial position of the center of the
magnetic trap in Ref.~\cite{Smerzi_02}.

\begin{figure}[t] 
\begin{center} 
\includegraphics[scale=1.2]{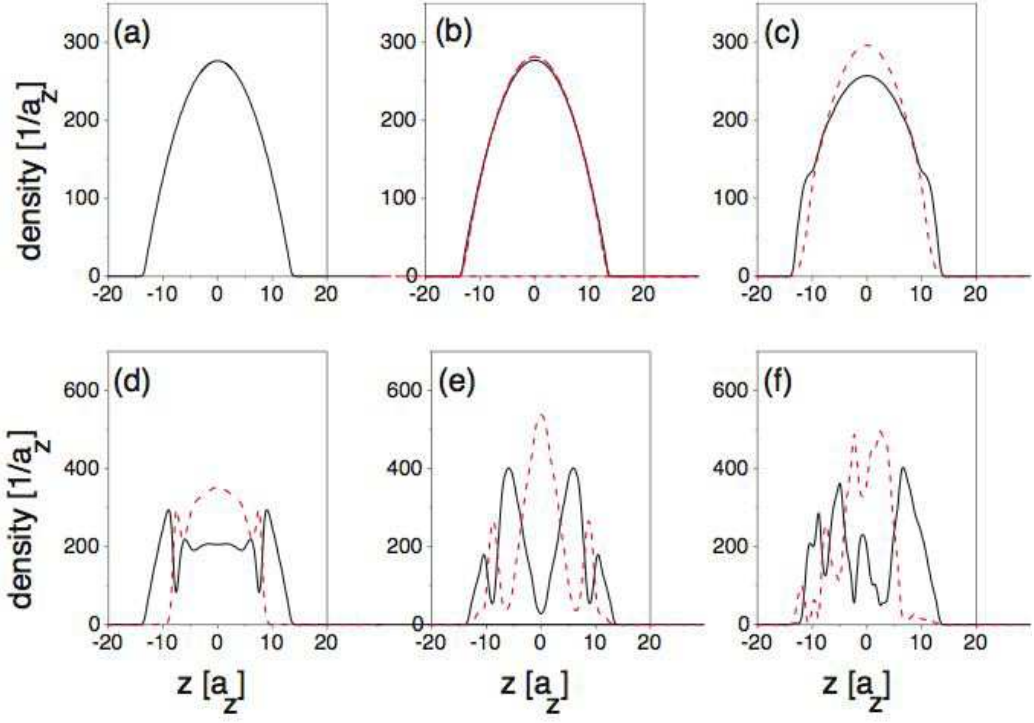} 
\caption{(color online) Condensate density of ground state (solid
curve) and excited state (dashed curve) components for the interstate
two-particle correlation parameter at zero separation
$\alpha^{(2)}_{12} = 1$, after the first $\pi/2$ pulse for a sequence
of inter-pulse times (a) $T=0$ s, (b) $T=0.36$ s, (c) $T=1$ s, (d)
$T=2$ s, (e) $T=50$ s and (f) $T=220$ s.  At $T=0$ s the two spin
components have the same amplitude since the microwave pulse is very
short but dephased of $\pi/2$ for a $\pi/2$ pulse.  }
\label{density_position_corr1} 
\end{center} 
\end{figure}

The dynamics is different for $\alpha^{(2)}_{12}=1$.
Fig.~\ref{density_position_corr1} shows the evolution of the position
dependent density of the two atomic states for a sequence of
inter-pulse times $T=0$ s (Fig.~\ref{density_position_corr1}(a)),
$T=0.36$ s (Fig.~\ref{density_position_corr1}(b)), $T=1$ s
(Fig.~\ref{density_position_corr1}(c)), $T=2$ s
(Fig.~\ref{density_position_corr1}(d)), $T=50$ s
(Fig.~\ref{density_position_corr1}(e)), and $T=220$ s
(Fig.~\ref{density_position_corr1}(f)).  The phase separation appears
later, around $0.4$ s, and evolves more slowly than the case of strong
interstate interaction strength ($\alpha^{(2)}_{12}=2$).  Since the
interaction strength of the ground state is larger than in the excited
state, the ground state density protrudes beyond the excited state
density ($T=1$ s), and at later times the phase separation evolves in
a complex way under the effect of the instability ($T=2$ s and $50$ s)
and eventually a symmetry breaking occurs ($T=220$ s).

A rough estimate of the time scale at which the system becomes 
sensitive to the phase-separation instability, $\tau_{\text{ps}}$, can 
be obtained by using the expression derived by Timmermans in 
Ref.~\cite{Timmermans_98} for a homogeneous system: 
\begin{equation} 
\tau_{\text{ps}} = 2\pi/ |\Omega_{-,k_f}| = 2\pi \hbar /m|c_{-}|^2 
\,, 
\end{equation} 
where 
\begin{equation} 
\Omega_{-,k}^2=c_{-}^2k^2+(\hbar k^2/2m)^2
\label{dispersion_relation}
\end{equation} 
is the dispersion of the double condensate excitation, and the 
parameter 
\begin{equation} 
c_{-}^2 = \frac{\hbar^2}{m a_{\perp}^2} \left(
\alpha^{(2)}_{11}a_{11}n_1+ \alpha^{(2)}_{22} a_{22} n_2
-\sqrt{(\alpha^{(2)}_{11} a_{11}n_1)^2+(\alpha^{(2)}_{22} a_{22}
n_2)^2 + (2(\alpha^{(2)}_{12}a_{12})^2 - \alpha^{(2)}_{11}
\alpha^{(2)}_{22} a_{11}a_{22})2 n_1 n_2} \right)
\label{sound_velocity}
\end{equation}
is the phonon-like sound velocity at low momenta.  For instability,
$c_{-}^2 < 0$ and $\Omega_{-,k_f}^2 < 0$.  The fastest growing mode
has wave number $k_f=\sqrt{2}m|c_{-}|/\hbar$, and grows with an
initial rate of $m|c_{-}|^2/\hbar$.
Eqs.~(\ref{dispersion_relation})-(\ref{sound_velocity}) show that the
local phase separation of the two condensate components and the
symmetry breaking are due to the cross interaction terms under the
condition $c_{-}^2 < 0$ which occurs when $(\alpha^{(2)}_{12}a_{12})^2
> \alpha^{(2)}_{11}\alpha^{(2)}_{22} a_{11}a_{22}/2 $.
Eq.~(\ref{sound_velocity}) shows that the time at which the symmetry
breaking starts depends on the atomic density; the higher the density
(corresponding to larger parameter $|c_{-}|$), the smaller the
symmetry breaking appearance time.  We obtain a value of
$\tau_{\text{ps}} = 35$ ms which is smaller than the time from which
the phase separation starts in Fig.~\ref{density_position}.  This
discrepancy is probably due to the space-dependent atomic density that
locally changes the value of $\tau_{\text{ps}}$, increasing from the
center of the clouds where the density is higher to the edge where the
density is smaller.  As the instability depends on the density of
atoms, the local spiked structures start to develop at the center of
the trap and then spread throughout the clouds.

\begin{figure}[t] 
\begin{center} 
\includegraphics[scale=1.2]{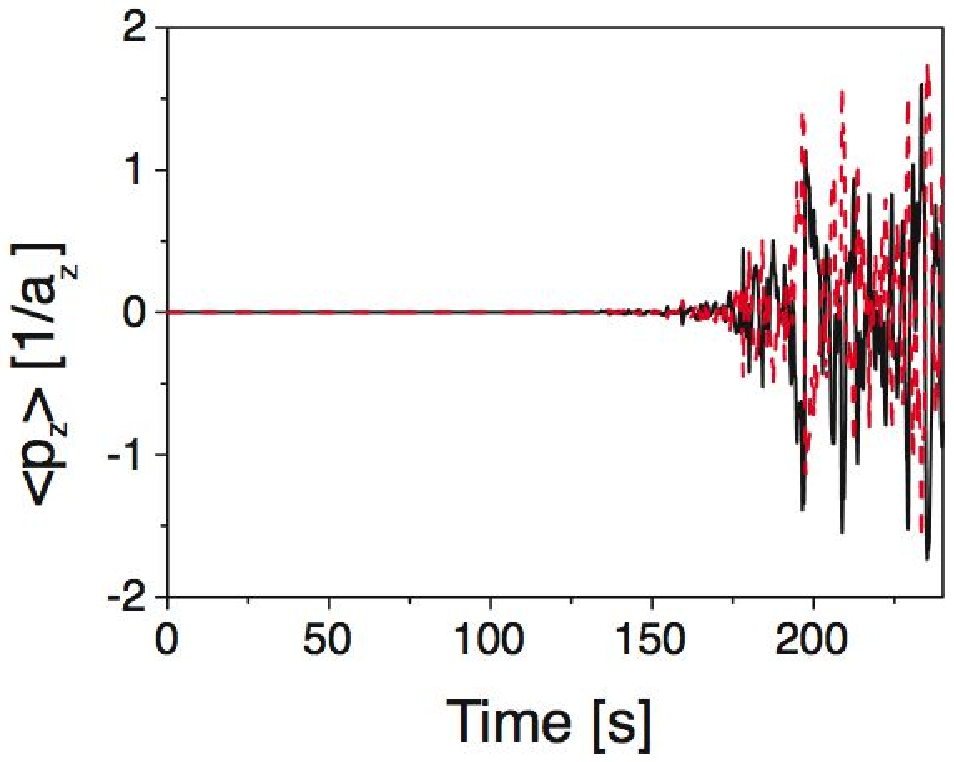} 
\caption{(color online) Evolution of the mean value of the momentum of
ground (solid curve) and excited (dashed curve) state component after
the first $\pi/2$ pulse as a function of time for
$\alpha^{(2)}_{12}=1$.  }
\label{mean_momentum_corr1}
\end{center} 
\end{figure}

Note that the case of $\alpha^{(2)}_{12} = 1$ with the same numerical
parameters values yields $\tau_{\text{ps}} \approx 7$ s and the
numerical calculation yields a value beyond $120$ s for the appearance
of the symmetry breaking (Fig.~\ref{mean_momentum_corr1}).  In this
case, the phase separation appears much earlier, $\approx 0.4$ s
before the symmetry breaking takes place.  Indeed as the intrastate
and interstate interaction strengths are almost identical, the BEC
clouds are not strongly perturbed after the first $\pi/2$ pulse.
Hence, the two BEC components are barely unstable and phase separation
and the symmetry breaking does not occur until much later than in the
case of $\alpha^{(2)}_{12}=2$.  This clearly increases the stability
of the clock.  The smaller the difference between the interstate and
intrastate interaction strength, the longer the interrogation time and
the higher the stability of the clock.

\begin{figure}[t] 
\begin{center} 
\includegraphics[scale=1.2]{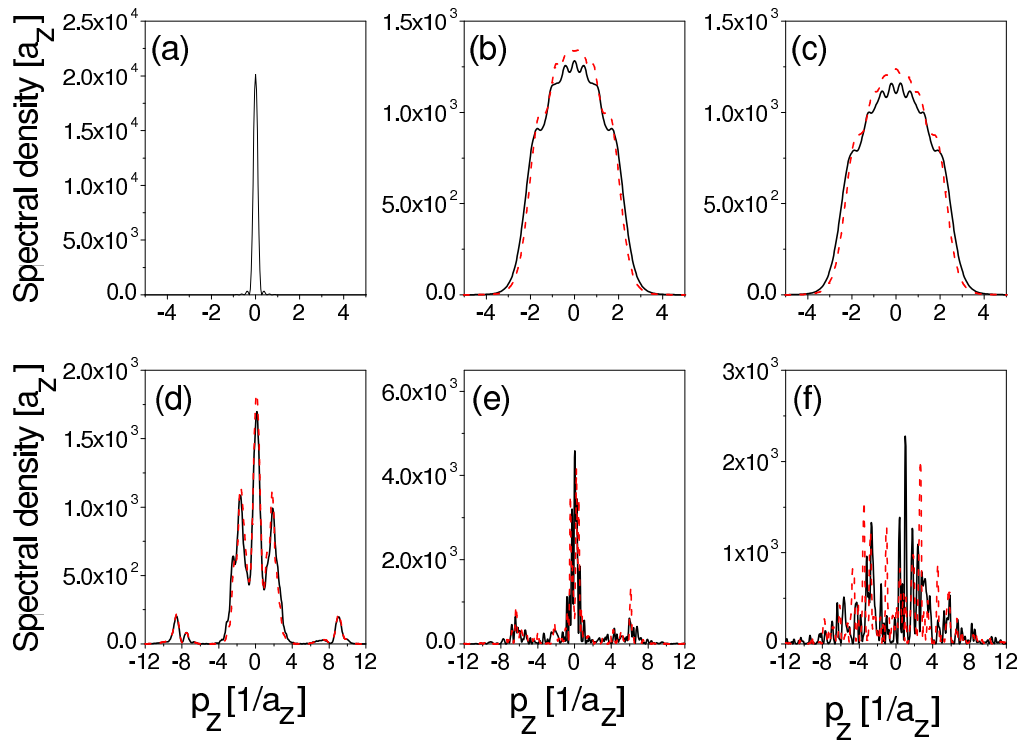} 
\caption{(color online) Condensate momentum density of ground (solid
curve) and excited (dashed curve) state component for the interstate
two-particle correlation parameter at zero separation
$\alpha^{(2)}_{12} = 2$, after the first $\pi/2$ pulse for a sequence
of times (a) $T=0$ s, (b) $T=0.18$ s, (c) $T=0.22$ s, (d) $T=0.26$ s,
(e) $T=0.5$ s and (f) $T=1$ s.  }
\label{density_momentum} 
\end{center} 
\end{figure}

\begin{figure}[t] 
\begin{center} 
\includegraphics[scale=1.2]{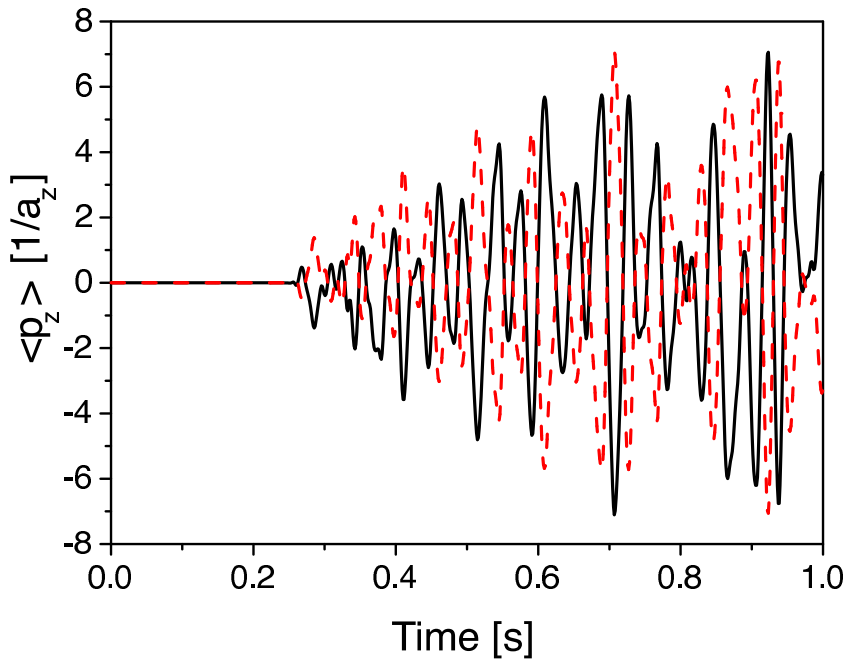} 
\caption{(color online) Evolution of the mean value of the momentum of
ground (solid curve) and excited (dashed curve) state component after
the first $\pi/2$ pulse as a function of time for $\alpha^{(2)}_{12} =
2$.  }
\label{mean_momentum}
\end{center} 
\end{figure}

The spiked structure and phase separation in position space is
correlated with a delocalization in momentum space created by the
strong excitation of the two BEC components as the dynamics proceeds.
Indeed the calculated density of the atoms in momentum space is
completely delocalized beyond $0.5$ s, as shown in
Fig.~\ref{density_momentum}.  The increased width of the momentum
distribution observed from $t=0$ s to $t=0.22$ s is due to the fact
that after the first $\pi/2$ pulse the interaction energy converts to
the kinetic energy.  To monitor the symmetry breaking shown in
Fig.~\ref{phase}(d), we calculate the mean value of the axial momentum
of each spin component as a function of time, $\langle p_{i_{z}}(t)
\rangle = \int p_{i_{z}}(t)\big| \psi_{i}(p_{z},t)\big|^2dp_{z}$ where
$\psi_{i}(p_{z},t)$ is the Fourier transform of the condensate wave
function $\psi_{i}(z,t)$.  Fig.~\ref{mean_momentum} shows the
evolution of the mean value of momentum of the two condensates as a
function of time.  The mean value of the momentum for each spin
component is zero from $0$ to $0.24$ s, and then the mean momentum of
each component starts oscillating in time, but the total momentum is
conserved.  We note that the symmetry breaking appears after the
density starts showing an irregular profile of small amplitude at its
center.  Comparison of Fig.~\ref{mean_momentum} and
Fig.~\ref{mean_momentum_corr1} shows that phase separation occurs at
much later times.
 
\begin{figure}[t] 
\begin{center} 
\includegraphics[scale=1.2]{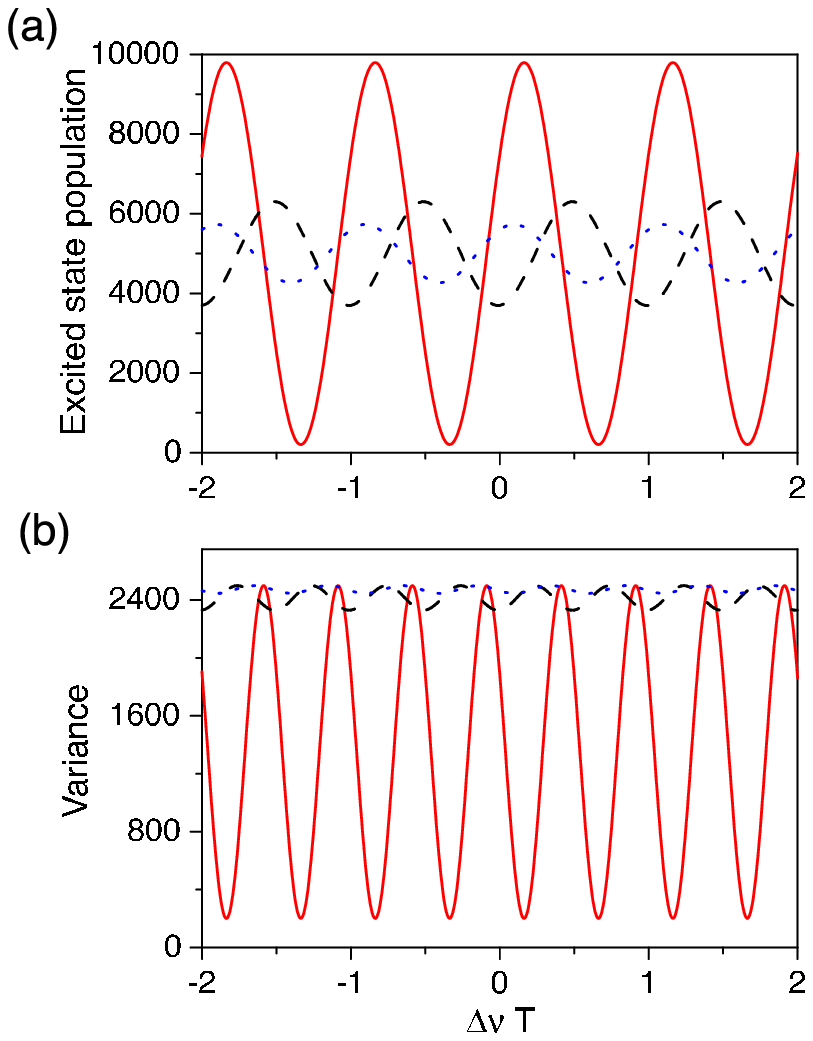} 
\caption{(color online) Excited state population (a) and it variance
(b) as a function of $\Delta \nu T$, for a sequence of inter-pulse
times: $T=0.18$ s (solid curve), $T=0.5$ s (dashed curve) and $T=1$ s
(dotted curve) for $\alpha^{(2)}_{12} = 2$.  }
\label{prob_var} 
\end{center} 
\end{figure}

We now analyze the effect of the mean-field dynamics of the system on 
the performance of the clock.  Fig.~\ref{prob_var} shows the 
calculated probability $P_e$ for finding atoms in the excited state 
immediately after the second $\pi/2$ Ramsey pulse as a function of the 
detuning $\Delta \nu$ of the microwave frequency from the atomic 
transition frequency times the interrogation time, $\Delta \nu T$.  
The curves in Fig.~\ref{prob_var}(a) have been calculated for three 
values of the interrogation time, $T=0.18$ s (solid curve), $0.5$ s 
(dashed curve), and $1$ s (dotted curve).  The fringe contrast 
decreases as the interrogation time increases.  Fig.~\ref{prob_var}(b) 
plots the variance of the excited population $NP_e(1-P_e)$ 
\cite{Itano_93, Wineland_94}.  The variance is large at time $T = 0.5$ 
and 1.0 s where the condensate profiles are spiked and asymmetrical, 
and this results in poor stability of the clock.  The interrogation 
time of $T=0.18$ s where the condensates show a smooth and symmetrical 
profile lead to a frequency stability of $2.6 \times 
10^{-12}\sqrt{T_{c}/\tau}$ where $T_{c}$ is the cycle period and 
$\tau$ the averaging time.  It is important to note that the stability 
can be improved by further decreasing the axial and radial 
frequencies, keeping a high anisotropy ratio so that the quasi-1D 
regime remains.  The improvement will depend on how low the axial 
frequency can be made without causing fluctuations of the trapping 
magnetic field.  The goal is to further lower the density so as to 
increase the time at which the phase separation of the two spin 
components after the first microwave pulse, and hence increase the 
interrogation time.  The total number of atoms can also be 
independently optimized.  Thus, it will hopefully be possible to reach 
a stability beyond $10^{-12}\sqrt{T_{c}/\tau}$.

We can express the Allan standard deviation for a Ramsey fringe 
experiment as $\sigma = \frac{1}{\pi \nu_0 T \sqrt{N}} 
\sqrt{T_{c}/\tau}$, as a function of the axial trapping frequency.  
The interrogation time is fixed by the density of the atoms; the 
higher the density, the smaller the interrogation time.  By fixing the 
interrogation time $T_{\mathrm{fix}}$ to correspond to a given density 
of atoms ($n=N/l_z$, where $l_z=(3g_{11}^{1D}N/2m\omega_{z}^2)^{1/3}$ 
it is possible to vary the number of atoms and the axial frequency.  
By equating the density calculated for different number of atoms and 
trap frequency, we derive the number of the atoms as function of the 
axial trapping frequency: 
\begin{equation} 
N=N_{\mathrm{fix}}\frac{ \omega_{z,\mathrm{fix}}}{\omega_z} 
\end{equation} 
where $N_{\mathrm{fix}}$ and $\omega_{z,\mathrm{fix}}$ are 
respectively the given number of atoms and axial trapping frequency 
that fix the density and the interrogation time $T_{\mathrm{fix}}$.  
Then the Allan standard deviation is: 
\begin{equation} 
\sigma(\omega_z)=\frac{1}{\pi \nu_0 T_{\mathrm{fix}}
\sqrt{N_{\mathrm{fix}}}} \Big(
\frac{\omega_{z}}{\omega_{z,\mathrm{fix}}} \Big)^{1/2}
\sqrt{T_{c}/\tau} \,,
\end{equation} 
and varies as the square root of the axial trapping frequency.  In our
case, using $N_{\mathrm{fix}} = 10^4$ atoms and
$\omega_{z,\mathrm{fix}} = 0.5$ Hz, we find $T_{\mathrm{fix}} = 180$
ms for $\alpha^{(2)}_{12}=2$, and the new values of the interrogation
time obtained by varying the axial trapping frequency are within about
$10$\% of $T_{\mathrm{fix}}$.

The projection noise does not have a simple expression as a function 
of the radial frequency because the instability of the BEC depends 
both on the density and the geometry of the cloud.  Furthermore, the 
variation of the radial frequency deeply affects the geometry of the 
cloud and changes the interaction strength $g_{11}^{1D}$.  

Normally we should add to the projection noise, the frequency noise 
due to the fluctuation of the density introduced by the fluctuation of 
the total number of atoms.  However, as we shall see in the next 
section, there is a simple method to cancel the collisional frequency 
shift and then the noise due to density fluctuations will cancel too.  

\section{Improvement of the Clock: Cancellation of the Collisional 
Shift}  \label{improvement}

In this section we analyze the possibility of improving the clock by 
cancelling the collisional shift.  One of the advantages resulting 
from the cancellation of the collisional shift is that the clock 
becomes insensitive to the variation of the total number of atoms.  
Indeed, even if the density is low, nevertheless 
the variation of the total number of atoms at each cycle period 
creates a variation of the density of the atoms and this introduces 
noise that limits the stability of the clock.  As the collisional 
frequency shift depends on the density of atoms, the idea is to cancel 
the collisional shift so that such a variation of the density of atoms 
does not affect the stability of the clock.  Note that the variation 
of the number of atoms we are referring to is not due to quantum 
fluctuations but rather due to experimental fluctuations in the number 
of atoms.  Here we present two different proposals to overcome this 
problem.  

\begin{figure}[t] 
\begin{center} 
\includegraphics[scale=1.2]{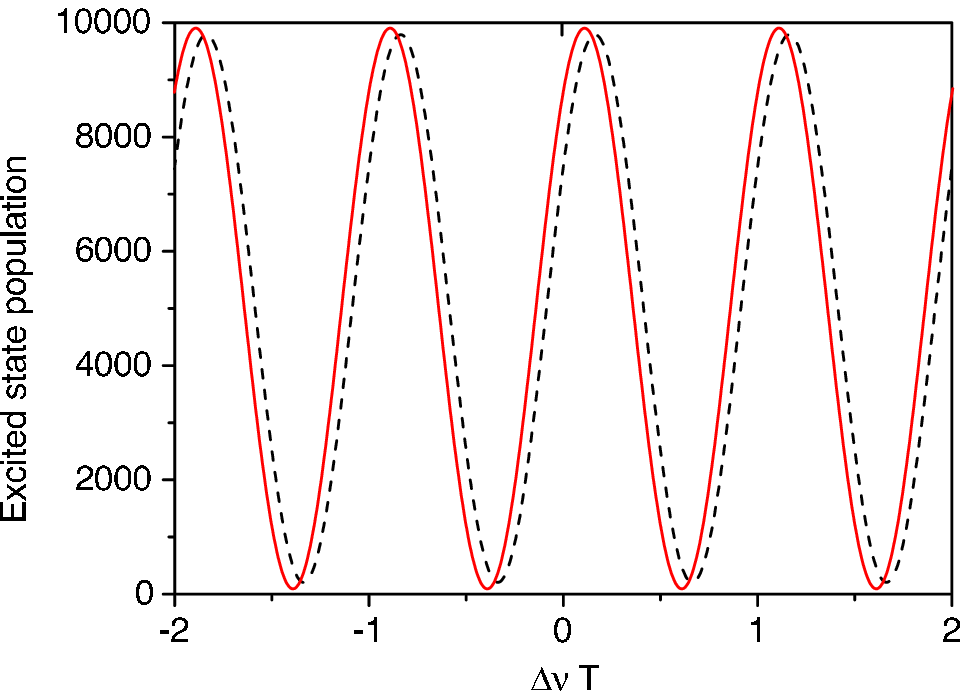} 
\caption{(color online) Excited state population as a function of the
detuning $\Delta \nu$ of the microwave pulse from atomic transition
times the interrogation time in the presence (dashed curve) or without
collisional (solid curve) for an inter-pulse time $T=0.18$ s, and the
interstate two-particle correlation at zero separation
$\alpha^{(2)}_{12} = 2$.  }
\label{cancel_collision} 
\end{center} 
\end{figure}

A small density of atoms in the excited state minimizes the
collisional dipolar relaxation loss.  But the clock sensitivity to
quantum fluctuations increases if the population of atoms in the
excited state is too small.  So an optimization of the density of
atoms is necessary to obtain both good signal-to-noise ratio and a
long clock time, yet having small collisional dipolar relaxation.  We
can use the Zeeman shift to compensate the collisional shift.  This
method can be applied both to a normal cold atomic cloud and a BEC,
and has been proposed in Ref.~\cite{Harber_02}.  However, this method
requires adjusting two parameters and seems not to be simple to
implement.  We will not use it in this paper.  Instead, we study the
cancellation of the collisional shift terms by playing them off
against each other as discussed by Gibble and Verhaar in
Ref.~\cite{Gibble_95} for a thermal cesium atom clock.  The expression
for the collisional frequency shift is \cite{Harber_02,Gibble_95,
Koelman_88, Oktel_02,Naraschewski_99,Tiesinga_92}
\begin{equation} 
\Delta \nu_{\mathrm{int}} = \frac{\hbar}{m\pi a_{\perp}^2}
(\alpha^{(2)}_{12} a_{12} n_1 + \alpha^{(2)}_{22}a_{22}n_2-
\alpha^{(2)}_{11}a_{11}n_1 - \alpha^{(2)}_{12} a_{12}n_2) \,. 
\end{equation}
where $n_1$ and $n_2$ are the density of atoms per unit length in the 
ground and excited state respectively.  By equating $\Delta 
\nu_{\mathrm{int}}$ to zero, we obtain a simple relation between the 
density of the atoms in the two states: 
\begin{equation} 
\frac{n_2}{n_1} = \frac{ \alpha^{(2)}_{12} a_{12} - \alpha^{(2)}_{11}
a_{11}} {\alpha^{(2)}_{12}a_{12} - \alpha^{(2)}_{22}a_{22} } \,.
\label{n2n1ratio}
\end{equation}
The above relation is satisfied only if $\alpha^{(2)}_{12} >
\alpha^{(2)}_{11}a_{11}/a_{12}$ and $\alpha^{(2)}_{12} >
\alpha^{(2)}_{22}a_{22}/a_{12}$, or $\alpha^{(2)}_{12} <
\alpha^{(2)}_{11}a_{11}/a_{12} $ and $\alpha^{(2)}_{12} <
\alpha^{(2)}_{22}a_{22}/a_{12}$.  When $\alpha^{(2)}_{11} =
\alpha^{(2)}_{22} = 1$, and $a_{12}$ is only slightly different from
$a_{11}$ and $a_{22}$, such that $a_{2} < a_{12} < a_{11}$, the
cancellation of the collisional shift is possible only for
$\alpha^{(2)}_{12} \geq 1.02$ or $\alpha^{(2)}_{12} \leq 0.97$.  For
$\alpha^{(2)}_{12} =1$ the ratio $n_2/n_1$ in Eq.  (\ref{n2n1ratio})
is negative, and therefore cancellation of the collisional shift is
not possible.  For the scattering lengths of $^{87}$Rb and
$\alpha^{(2)}_{12} =2$, the density ratio is $n_2 =0.95\, n_1$.  Such
a ratio should maintain the sizes of the two components to be almost
identical.  For a ratio of densities close to unity, the Zeeman
frequency shift terms due to the size of the atomic clouds will be
small.  However, a problem originates from $\ket{2}$-$\ket{2}$
collisional dipolar relaxation which can be important at high density
and this can limit the clock run-time.  But in a quasi-1D system, the
inelastic ultra-cold collisions are further reduced.  To illustrate
the improvement of collisional shift, Fig.~\ref{cancel_collision}
shows the calculated excited state population as a function of $\Delta
\nu T$ for an inter-pulse time $T=0.18$ s.  The dashed curve is for
the case of $\pi/2$ pulses and the solid curve shows the case of the
slightly different populations suggested above.  In the latter case,
the collisional shift is cancelled and this gives a better fringe
contrast and a smaller frequency shift.  The improvement of fringe
amplitude is due to the fact that the difference in the phase of the
condensates is space-independent as the collisional shift is
cancelled.

The remaining frequency shift is now due to kinetic and Zeeman shift.  
The frequency shift due to the kinetic term, 
\begin{equation} 
\Delta \nu_{\mathrm{kinetic}} = \frac{ \langle p_{z}^{2}\rangle_{2} 
- \langle p_{z}^{2}\rangle _{1}}{4\pi m\hbar} \,, \label{dnu_kin} 
\end{equation} 
is $-10.997$ Hz and the shift due to the Zeeman terms, 
\begin{equation} 
\Delta \nu_{\mathrm{Zeeman}} = \frac{\langle m 
\omega_{z}^2z^{2}\rangle_{2} - \langle m \omega_{z}^2z^{2}\rangle_{1}}
{4\pi \hbar} \,, \label{dnu_mag} 
\end{equation} 
is $-0.266$ Hz, giving a total frequency shift of $\Delta \nu =
-11.263$ Hz, in good agreement with the frequency shift $11.12$ Hz 
determined from the interference pattern of the solid curve in 
Fig.~\ref{cancel_collision}.  In Eqs.~(\ref{dnu_kin}-\ref{dnu_mag}), 
the symbol $\langle \ldots \rangle_{i}$ denotes the expectation value 
calculated with the wave function $\psi_{i}(t=T)$.  We see that the 
shift introduced by the Zeeman terms is small because it is 
proportional to the difference of the square of the size of each 
condensate and this difference is small.  The shift due to the kinetic 
terms is proportional to the difference of the square of the width of 
the spectral density of each condensate.  As the size of the 
condensates and the width of their spectral density change in time, 
the shifts are time dependent.  

\section{Conclusion}  \label{Conclusion} 

We modeled a microwave frequency atomic clock using a configuration of
BEC atoms in a highly elongated magnetic trap.  We showed that the
stability of the clock for a trap radial frequency $\omega_r/2\pi=120$
Hz and axial frequency $\omega_z/2\pi=0.5$ Hz is $2.6\times
10^{-12}\sqrt{T_{c}/\tau}$ and $1\times 10^{-12}\sqrt{T_{c}/\tau}$ for
$\alpha^{(2)}_{12}=2$ and $\alpha^{(2)}_{12}=1$ respectively.  The
performance of the clock is related to the configuration of the trap
and can be improved by running the clock with an even weaker axial
trapping frequency.  We found a dynamical instability that results in
phase separation and limits the clock stability and accuracy if a long
interrogation time is used depending on the density of atoms and on
the ratio between the intrastate and interstate two-particle
correlation parameter at zero separation.  We considered optimization
of the experimental parameters maximize the stability and accuracy of
the clock.  For a $^{87}$Rb BEC, the collisional shift terms can be
cancelled by playing them off against each other by adjusting the
population in the ground and excited states with the Ramsey pulses so
that the problem of density fluctuations does not strongly affect the
stability and the accuracy of the clock.  This can be done only if the
two-particle correlation parameters satisfy the relations
$\alpha^{(2)}_{12} > \alpha^{(2)}_{11}a_{11}/a_{12}$ and
$\alpha^{(2)}_{12} > \alpha^{(2)}_{22}a_{22}/a_{12}$, or
$\alpha^{(2)}_{12} < \alpha^{(2)}_{11}a_{11}/a_{12}$ and
$\alpha^{(2)}_{12} < \alpha^{(2)}_{22}a_{22}/a_{12}$.  The equality of
these previous relations automatically cancel the collisional shift.

\begin{acknowledgments}

We thank Drs. R. Tasgal and Y. Japha for useful conversations.  
This work was supported in part by grants from the U.S.-Israel 
Binational Science Foundation (grant No.~2002147), the Israel Science 
Foundation for a Center of Excellence (grant No.~8006/03), and the 
German Federal Ministry of Education and Research (BMBF) through the 
DIP project. 
\end{acknowledgments}


\begin{thebibliography}{99}
\bibitem{good_q_n}  
Clearly, in a magnetic field, the only good quantum numbers are 
$m_{f}$.  In the presence of a magnetic field, what we mean by states 
$\ket{f,m_f}$ are the states $\ket{m_f}$ that correlate at zero field 
to $\ket{f,m_f}$.
\end{thebibliography}
\end{document}